\documentclass{wsprocs}
\RequirePackage{xspace}
\usepackage{graphicx}
\usepackage{dcolumn}
\usepackage{amsmath}
\usepackage{epsfig}
\usepackage{multirow}
\usepackage{relsize}
\def\qqbar {\ensuremath{q\overline q}\xspace}
\def\babar{\mbox{\slshape B\kern-0.1em{\smaller A}\kern-0.1em
    B\kern-0.1em{\smaller A\kern-0.2em R}}}
\def\Abar    {\kern 0.18em\overline{\kern -0.18em A}{}\xspace}
\def\Kbar    {\kern 0.18em\overline{\kern -0.18em K}{}\xspace}
\def\Bbar    {\kern 0.18em\overline{\kern -0.18em B}{}\xspace}

\def\BB      {\ensuremath{B\Bbar}\xspace} 
\def\Bz      {\ensuremath{B^0}\xspace}
\def\Bzb     {\ensuremath{\Bbar^0}\xspace}
\def\BzBzb   {\ensuremath{\Bz {\kern -0.16em \Bzb}}\xspace}
\def\Bu      {\ensuremath{B^+}\xspace}
\def\Bub     {\ensuremath{B^-}\xspace}

\def\BpBm    {\ensuremath{\Bu {\kern -0.16em \Bub}}\xspace}

\newcommand{\optbar}[1]{\shortstack{{\tiny (\rule[.4ex]{1em}{.1mm})}
  \\ [-.7ex] $#1$}}
\def\BorBbar    {\kern 0.18em\optbar{\kern -0.18em B}{}\xspace}
\def\DorDbar    {\kern 0.18em\optbar{\kern -0.18em D}{}\xspace}
\def\KorKbar    {\kern 0.18em\optbar{\kern -0.18em K}{}\xspace}

\def\pep2{PEP-II}
\mathchardef\Upsilon="7107
\def\Y#1S{\ensuremath{\Upsilon{(#1S)}}\xspace}% no space before {...}!

\def\FourS {\Y4S}

\begin{document}

\title{\boldmath Polarization puzzle in $B\to\phi K^*$ 
and other $B\to VV$ at $\babar$}

\author{A. Gritsan}

\address{
LBNL, 1 Cyclotron Rd., M.S. 50A-2160, Berkeley, CA 94720, USA\\
E-mail: AVGritsan@lbl.gov}

\twocolumn[\maketitle\abstract{
With a sample of about 227 million $\BB$ pairs recorded 
with the $\babar$ detector we perform a full angular 
analysis of the decay $B^0\to\phi K^{*0}(892)$. 
Ten measurements include polarization, phases, and
five $C\!P$ asymmetries. We also observe the decay
$B^0\to\phi K^{*0}(1430)$. Polarization measurements
are performed with the $B\to\rho K^{*}(892)$, 
$B\to\rho\rho$, and $B\to\rho\omega$ decay modes, and
limits are set on the $B\to\omega K^{*}(892)$ decay rates.
These measurements help to understand the puzzle of
large fraction of transverse polarization observed in
$B\to\phi K^{*}$ decays and allow for new ways to 
study $C\!P$ violation and potential new amplitude 
contributions.}]

\section{Introduction}%1

The decay $B\to\phi K^*(892)$ is expected to have contributions
from $b\to s$ loop penguin transitions while the 
tree-level transition is suppressed in the Standard Model. 
Angular correlation 
measurements and asymmetries are particularly sensitive to 
amplitudes arising outside the Standard Model~\cite{bvv1}. 
The first evidence for this decay was provided by 
the CLEO~\cite{cleo:phikst} and \babar~\cite{babar:phikst} 
experiments. The large fraction of transverse polarization
observed by \babar~\cite{babar:vv} and confirmed by 
BELLE~\cite{belle:phikst} was a surprise and enabled 
a full angular analysis described by ten parameters for
contributing amplitudes and their phases.

Similarly, the decays $B\to\rho K^*(892)$ and 
$B\to\omega K^*(892)$ are expected to have contributions 
from $b\to s$ loop transitions with some tree contributions. 
Polarization measurements in these channels may help 
in understanding the $B\to\phi K^*$ polarization puzzle.
The decays $B\to\rho\rho$ and $\omega\rho$ are expected to 
proceed through the tree-level $b\to u$ transition and through
CKM-suppressed $b\to d$ penguin transitions.
These are particularly interesting modes for the CKM angle $\alpha$
studies and have the advantage of a larger decay rate and smaller 
uncertainty in penguin pollution compared to $B\to\pi\pi$.
The \babar~\cite{babar:vv,babar:rhoplrhomn} and 
the BELLE~\cite{belle:rhozrrhopl} 
experiments reported observation of the 
$B\to{\rho K^{*}}$ and ${\rho\rho}$ decays. 

The angular distribution of the $B\to VV$ decay products 
are expressed as a function of $\cos\theta_i$ and $\Phi$, 
where $\theta_i$ is the helicity angle of a $\phi$, $K^*$, 
$\rho$, or $\omega$, and $\Phi$ is the angle between the 
two resonance decay planes. The differential decay width 
has three complex amplitudes $A_\lambda$ corresponding to 
the vector meson helicity $\lambda=0$ or 
$\pm 1$~\cite{bvv1,bvv2}. The last two can be expressed 
in terms of $A_{\parallel}=(A_{+1}+A_{-1})/\sqrt{2}$ and
$A_{\perp}=(A_{+1}-A_{-1})/\sqrt{2}$.

In this paper we present the latest results from $\babar$
in a number of charmless $B\to VV$ decays.
We measure the branching fraction, the polarization 
parameters $f_L={|A_0|^2/\Sigma|A_\lambda|^2}$,
$f_{\perp}={|A_{\perp}|^2/\Sigma|A_\lambda|^2}$,
and the relative phases
$\phi_{\parallel} = {\rm arg}(A_{\parallel}/A_0)$,
$\phi_{\perp} = {\rm arg}(A_{\perp}/A_0)$.
We allow for $C\!P$-violating differences between
the $\Bbar^0$ ($Q=+1$) and ${B}^0$ ($Q=-1$) decay 
amplitudes ($\Abar_{\lambda}$ and $A_{\lambda}$),
and derive vector triple-product asymmetries~\cite{bvv1}: 
%%%%%%%%%%%%%%%%%%%%%%%
\begin{eqnarray}
\label{eq:tripleprod}
{\cal A}_T^{\parallel,0}= {1\over 2}\left(
{ {\rm Im}(A_{\perp}A^{*}_{\parallel,0}) \over \Sigma|A_\lambda|^2 } +
{ {\rm Im}(\Abar_{\perp}\Abar^{*}_{\parallel,0}) \over \Sigma|\Abar_\lambda|^2 }\right)
\nonumber 
\end{eqnarray}
%%%%%%%%%%%%%%%%%%%%%%%

The $B$ flavor sign $Q$ can be determined in the self-tagging 
final state, then we have ten independent measured quantities:
%%%%%%%%%%%%%%%%%%%%%%%
\begin{eqnarray}
\label{eq:observe}
\nonumber 
n_{\rm sig}^Q={n_{\rm sig}}~\!(1+Q~\!{{\cal A}_{C\!P}})/2;\\
\nonumber 
f_{L}^{~\!Q} = f_{L}~\!(1+Q~\!{\cal A}_{C\!P}^{0}); \\
\nonumber 
f_{\perp}^{~\!Q} = f_{\perp}~\!(1+Q~\!{\cal A}_{C\!P}^{\perp});\\
\nonumber 
\phi_{\parallel}^Q = \phi_{\parallel}+Q~\!\Delta \phi_{\parallel};\\
\nonumber 
\phi_{\perp}^Q = \phi_{\perp}+{\pi\over 2}
 + Q~\!(\Delta \phi_{\perp}+{\pi\over 2}).
\nonumber
\end{eqnarray}
%%%%%%%%%%%%%%%%%%%%%%%

\section*{Experimental technique}

We use data collected with the 
\babar\ detector~\cite{babar} at the \pep2 asymmetric-energy 
$e^+e^-$ collider operated at the 
center-of-mass energy of the $\FourS$ resonance
($\sqrt{s}=10.58$~GeV). 
We fully reconstruct vector-vector $B$ meson decays 
involving $\phi$, $\rho$, $\omega$, and $K^*$ resonances.
We identify $B$ meson candidates using two variables:
$m_{\rm{ES}} = [{ (s/2 + \mathbf{p}_i \cdot \mathbf{p}_B)^2 / E_i^2 - 
\mathbf{p}_B^{\,2} }]^{1/2}$ and
$\Delta{E}=(E_iE_B-\mathbf{p}_i$$\cdot$$\mathbf{p}_B-s/2)/\sqrt{s}$,
where $(E_i,\mathbf{p}_i)$ is the initial state four-momentum,
and $(E_B,\mathbf{p}_B)$
is the four-momentum of the reconstructed $B$ candidate.
To reject the dominant quark-antiquark continuum background
we apply event-shape requirements.

We~use an unbinned maximum-likelihood (ML) fit to
extract signal parameters.
There are several event categories $j$: signal, 
continuum~$\qqbar$, combinatoric $\BB$ background, and specific
$B$-decay background modes.
The likelihood for each candidate~$i$ is defined as
${\cal L}_i$ $=$ $\sum_{j,k}n_{j}^k\, 
{\cal P}_{j}^k(\vec{x}_{i};\vec{\alpha};\vec{\beta})$,
where each of the 
${\cal P}_{j}^k(\vec{x}_{i};\vec{\alpha};\vec{\beta})$ is 
the probability density function for input variables.
The $n_{j}^k$ is the number of events with the $B$ flavor $k$
in the category~$j$.
The event yields $n_j$, asymmetries ${\cal A}_j$,
and the signal polarization parameters $\vec{\alpha}$
are obtained by maximizing 
${\cal L}=\exp(-\sum n_{j}^k)\prod{\cal L}_i$.

\begin{figure}[hbt]
\centerline{
\setlength{\epsfxsize}{1.0\linewidth}\leavevmode\epsfbox{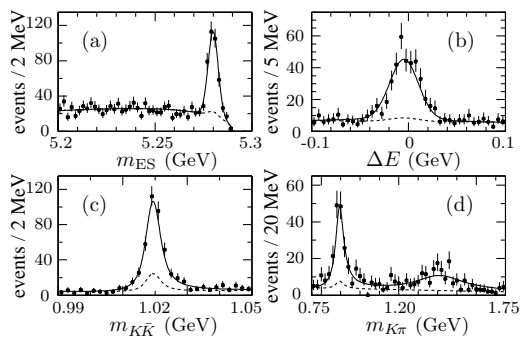}
}
\caption{\label{fig:projection1} 
Projections onto the variables 
$m_{\rm ES}$ (a), $\Delta E$ (b), $m_{K\!\Kbar}$ (c), and
$m_{K\!\pi}$ (d) for the signal $B^0\to\phi K^{*0}(892)$
and $\phi K^{*0}(1430)$ candidates combined.
}
\end{figure}

In Fig.~\ref{fig:projection1}
examples of fit input variables and
ML fit projections are shown, 
%%%%%%%%%%%%%%%%%%%%%%%%%%%%%%%%%%%%%%%%%%%%%%%%%%%%%%%%%
%%%%%%%%%%%%%%%%%%%%%%%%%%%%%%%%%%%%%%%%%%%%%%%%%%%%%%%%%
where data distributions are shown
with the signal enhanced with a requirement 
on the signal-to-background probability ratio 
calculated with the plotted variable excluded.

\section*{Results}

The results of our maximum likelihood fit to the sample
of $B^0\to\phi K^{*0}(892)$ candidates are summarized 
in Table~\ref{tab:results1}.
We observe, with more than 5$\sigma$ significance, 
non-zero contributions from all of the three amplitudes
$A_0$, $A_\perp$, and $A_\parallel$
($f_L+f_\perp+f_\parallel=1)$. We find 3$\sigma$ evidence 
for non-zero final-state-interaction phases 
(${\phi_\parallel}$ and ${\phi_\perp}$ differ from $\pi$
or zero). 
We also observe $B^0\to\phi K^{*0}(1430)$ decays.

In Table~\ref{tab:results2} we show results for all
$B\to VV$ modes with the dominant $b\to s$ penguin
contribution expected. Naive SU(3) decomposition of
%%%%%%%%%%%%%%%%%%%%%%%%%%%%%%%%%%%%%%%%%%%%%%%%
\begin{table}[h]%1
\caption{Summary of the $B^0\to\phi K^{*0}(892)$ fit results.
We show results for the ten primary signal fit parameters
and the derived branching fraction ${\cal B}$ and
triple-product asymmetries ${\cal A}_T^{\parallel}$
and ${\cal A}_T^{0}$.
}
\label{tab:results1}
\begin{center}
\begin{tabular}{|cc|}
\hline
\vspace{-3mm} & \\
   Fit parameter %%%%%%%%%%%%%%%%%%%%%%%%
 & Fit result 
\cr
\vspace{-3mm} & \\
\hline
\vspace{-3mm} & \\
  $n_{\rm sig}$ (events) %%%%%%%%%%%%%%%%%%%%%%%%
 & $201\pm 20\pm 6$  
\cr
\vspace{-3mm} & \\
  ${f_L}$  %%%%%%%%%%%%%%%%%%%%%%%%
 & $0.52\pm{0.05}\pm 0.02$  
\cr
\vspace{-3mm} &  \\
  ${f_\perp}$ %%%%%%%%%%%%%%%%%%%%%%%%
 & $0.22\pm{0.05}\pm 0.02$  
\cr
\vspace{-3mm} &  \\
  ${\phi_\parallel}$ (rad) %%%%%%%%%%%%%%%%%%%%%%%%
 & $2.34^{+0.23}_{-0.20}\pm 0.05$ 
\cr
\vspace{-3mm} &  \\
  ${\phi_\perp}$ (rad) %%%%%%%%%%%%%%%%%%%%%%%%
 & $2.47\pm 0.25\pm 0.05$    
\cr
\vspace{-3mm} &  \\
  ${\cal A}_{C\!P}$ %%%%%%%%%%%%%%%%%%%%%%%%
 & $-0.01\pm{0.09}\pm 0.02$  
\cr
\vspace{-3mm} &  \\
  ${\cal A}_{C\!P}^0$ %%%%%%%%%%%%%%%%%%%%%%%%
 & $-0.06\pm{0.10}\pm 0.01$ 
\cr
\vspace{-3mm} &  \\
  ${\cal A}_{C\!P}^{\perp}$ %%%%%%%%%%%%%%%%%%%%%%%%
 & $-0.10\pm 0.24\pm 0.05$ 
\cr
\vspace{-3mm} &  \\
  $\Delta \phi_{\parallel}$ (rad) %%%%%%%%%%%%%%%%%%%%%%%%
 & $0.27^{+0.20}_{-0.23}\pm 0.05$   
\cr
\vspace{-3mm} &  \\
  $\Delta \phi_{\perp}$ (rad) %%%%%%%%%%%%%%%%%%%%%%%%
 & $0.36\pm 0.25\pm 0.05$   
\cr
\vspace{-3mm} &  \\
\hline
\vspace{-3mm} &  \\
  ${\cal B}$ %%%%%%%%%%%%%%%%%%%%%%%%
 & $(9.2\pm{0.9}\pm 0.5)\times 10^{-6}$   
\cr
\vspace{-3mm} &  \\
  ${\cal A}_T^{\parallel}$ %%%%%%%%%%%%%%%%%%%%%%%%
 & $-0.02\pm{0.04}\pm 0.01$   
\cr
\vspace{-3mm} &  \\
  ${\cal A}_T^{0}$ %%%%%%%%%%%%%%%%%%%%%%%%
 & $+0.11\pm{0.05}\pm 0.01$   
\cr
%\vspace{-3mm} &  \\
\hline
\end{tabular}
\end{center}
\end{table}
%%%%%%%%%%%%%%%%%%%%%%%%%%%%%%%%%%%%%%%%%%%%%%%%
\begin{table*}[t]
\caption{The $\babar$ measurements of the branching fractions 
(${\cal B}$) and polarizations ($f_L$)
of the $B\to VV$ decays with the dominant
$b\to s$ penguin contribution. Relative coefficients
in front of the penguin, color-allowed and color-suppressed tree
amplitudes contributing to each decay mode are shown with 
{$\alpha_P$}, {$\alpha_T$}, and {$\alpha_C$}. 
Naive SU(3) decomposition is used for illustration.
The last column indicates the number of $\BB$ pairs used 
in each analysis. New preliminary results this year
are indicated by ``new'', while references are given to the
published results. The last error in the $\rho^+ K^{*0}$
channel has non-resonant decay rate uncertainty separated.
}
\label{tab:results2}
\begin{center}
\begin{tabular}{|c|ccc|c|c|c|} 
\hline
\vspace{-4mm} & & & & & &  \\
{$B$ decay} 
  & { $\alpha_P$} & { $\alpha_T$} & { $\alpha_C$}
  & {${\cal B} (10^{-6})$}  & $f_L$   & $N_{B\!\bar{B}} (10^{6})$  \cr
\vspace{-4mm} & & & & & &  \\
\hline
\vspace{-4mm} & & & & & &  \\
%%%%%%%%%%
{$\phi K^{*0}$} & { $\sqrt{2}$} &   {$0$} &  {$0$}  &
 {$9.2\pm 0.9$} { $\pm~{0.5}$}  & $0.52\pm 0.05\pm 0.02$ &  227 ({new})  \cr
{$\phi K^{*+}$} & { $\sqrt{2}$} &   {$0$} &  {$0$}  &
 {$12.7^{~+2.2}_{~-2.0}$} { $\pm~{1.1}$}  & $0.46\pm 0.12\pm 0.03$
 &  89 (publ.~\cite{babar:vv}) \cr
\vspace{-4mm} & & & & & &  \\
%%%%%%%%%%
\hline
%%%%%%%%%%
\vspace{-4mm} & & & & & &  \\
{$\rho^0 K^{*0}$} & { $1$} &   {$0$} &  { -$1$}  &
 --  & -- &  --  \cr
{$\rho^0 K^{*+}$} & { -$1$} &   { -$1$} & { -$1$}  &
 {$10.6^{~+3.0}_{~-2.6}$} { $\pm~{2.4}$}  &
  $0.96^{+0.04}_{-0.15}\pm 0.04$ 
 &  89 (publ.~\cite{babar:vv}) \cr
{$\rho^+ K^{*0}$} & { $\sqrt{2}$} &   {$0$} &  {$0$}  &
 {$\!\!17.0\pm${${2.9}$}}{$\pm$${2.0}^{+0.0}_{-1.9}\!\!$}  &
  $\!\!0.79$$\pm${$0.08$}$\pm${$0.04$}$\pm${$0.02\!\!$} & 
  89 ({new})  \cr
{$\rho^- K^{*+}$} & { -$\sqrt{2}$} &  { -$\sqrt{2}$} &  {$0$}  &
 {$<24$} (90\% C.L.)  & -- &  123 ({new})  \cr
\vspace{-4mm} & & & & & &  \\
%%%%%%%%%%
\hline
%%%%%%%%%%
\vspace{-4mm} & & & & & &  \\
{$\omega K^{*0}$} & { $1$} &  {$0$} & { $1$}  &
 {$<6.1$} (90\% C.L.)  & -- &  
 89 ({new}) \cr
{$\omega K^{*+}$} & { $1$} & { $1$} & { $1$}  &
 {$<7.4$} (90\% C.L.)  & -- &  
 89 ({new}) \cr
%%%%%%%%%%
\hline 
\end{tabular}
\end{center}
\end{table*}
%%%%%%%%%%%%%%%%%%%%%%%%%%%%%%%%%%%%%%%%%%%%%%%%
\begin{table*}[t]
\caption{The $\babar$ measurements of the branching fractions 
(${\cal B}$) and polarizations ($f_L$)
of the $B\to VV$ decays with the $b\to u$ tree and
$b\to d$ penguin contributions. Relative coefficients
in front of the color-allowed tree, color-suppressed tree, 
and penguin amplitudes 
are shown with {$\alpha_T$}, {$\alpha_C$}, and~{$\alpha_P$}. 
}
\label{tab:results3}
\begin{center}
\begin{tabular}{|c|ccc|c|c|c|} 
\hline
\vspace{-4mm} & & & & & &  \\
{$B$ decay} 
  & { $\alpha_T$} & { $\alpha_C$} & { $\alpha_P$}
  & {${\cal B}~(10^{-6})$}  & $f_L$   & $N_{B\!\bar{B}} (10^{6})$  \cr
\vspace{-4mm} & & & & & &  \\
\hline
\vspace{-4mm} & & & & & &  \\
%%%%%%%%%%
{$\rho^-\!\rho^+$} & { $\sqrt{2}$} &   {$0$} & { $\sqrt{2}$}  &
{$30\pm 4$} { $\pm~{5}$}  & $0.99\pm 0.03^{+0.04}_{-0.03}$
 &  89 (publ.~\cite{babar:rhoplrhomn}) \cr
%%%%%%%%%%
{$\rho^0\!\rho^+$} & { $1$} & { $1$} & {$0$} &
{$23^{~+6}_{~-5}$} { $\pm~{6}$}  & $0.97^{+0.03}_{-0.07}\pm 0.04$
 &  89 (publ.~\cite{babar:vv}) \cr
%%%%%%%%%%
{$\rho^0\!\rho^0$} &  {$0$} & {$1$}& { -$1$} &
  {$<1.1$} (90\% C.L.) &  -- &  227 ({new}) \cr
%%%%%%%%%%
\vspace{-4mm} & & & & & &  \\
\hline
\vspace{-4mm} & & & & & &  \\
%%%%%%%%%%
{$\omega\rho^+$} & { -$1$} & { -$1$} & { ${2}$}  &
{$12.6^{~+3.7}_{~-3.3}$} { $\pm~{1.8}$}  &
$0.88^{+0.12}_{-0.15}\pm 0.03$ &  89 ({new})   \cr
%%%%%%%%%%
{$\omega\rho^0$} & {$0$} & {$0$} & { -$\sqrt{2}$} &
  {$<3.3$} (90\% C.L.) &  -- & 89 ({new})  \cr
%%%%%%%%%%
\vspace{-4mm} & & & & & &  \\
\hline
\vspace{-4mm} & & & & & &  \\
%%%%%%%%%%
{$\phi\phi$} &  {$0$} & {$0$} & {$0$} &
  {$<1.5$} (90\% C.L.) &  -- & 89 ({new})  \cr
%%%%%%%%%%
\hline 
\end{tabular}
\end{center}
\end{table*}
%%%%%%%%%%%%%%%%%%%%%%%%%%%%%%%%%%%%%%%%%%%%%%%%
the relative penguin and tree diagrams is shown.
Transverse polarization fraction in both
$B\to\phi K^{*}(892)$ charge modes are close to $50\%$,
while this effect is less pronounced in the 
$B\to\rho K^{*}(892)$ modes. At the same time,
polarization measurements 
in the tree-dominated 
modes presented in Table~\ref{tab:results3} 
favor longitudinal polarization dominance.

For $B$ decays to light charmless particles
we expect the hierarchy of decay amplitudes to be 
$|A_0|\gg|A_{+1}|\gg|A_{-1}|$ 
under the assumption of either loop or tree diagram 
contribution~\cite{helicity}. 
Our measurements with the decay $B^0\to\phi K^{*0}(892)$
do not agree with the first inequality
but agree with the second one.
This suggests other contributions to the decay amplitude,
previously neglected, either within or beyond the 
Standard Model~\cite{bvv1,vvnew}.

We also observe the decays $B^0\to\phi K^{*0}(1430)$
which we find to be predominantly longitudinally polarized 
based on the $\phi$ helicity angle distribution.
The width and the angular distribution of the 
$K^{*0}(1430)$ resonance structure are not consistent with 
the pure $K_2^{*0}(1430)$ tensor state at more than 10$\sigma$. 
However, the angular distribution provides evidence of the 
longitudinally polarized tensor $K_2^{*0}(1430)$
contribution (with statistical significance of 3.2$\sigma$)
in addition to the scalar $K_0^{*0}(1430)$.
If the longitudinal polarization dominance holds for
the vector-tensor $B\to\phi K_2^{*}(1430)$ decays, 
this will point to the special role of the vector
current in the $B\to\phi K^{*}(892)$ polarization puzzle.

If one loop diagram dominates the $B\to\phi K^*$
decay amplitude, the direct ${C\!P}$ asymmetries 
${\cal A}_{C\!P}$, ${\cal A}_{C\!P}^{0}$, and 
${\cal A}_{C\!P}^{\perp}$, and the weak-phase differences 
$\Delta\phi_{\parallel}$ and $\Delta\phi_{\perp}$,
or alternatively ${\cal A}_T^{0}$ and 
${\cal A}_T^{\parallel}$, are expected to be negligible. 
These are interesting to look for new amplitude
contributions with different weak phases.

The rates of the $B^0\to\rho^+\rho^-$ and $B^+\to\rho^0\rho^+$ 
decays are larger than the corresponding rates of 
$B\to\pi\pi$ decays~\cite{pdg}. At the same time, 
the measurements of the $B\to\rho K^{*}$ branching 
fractions do not show significant enhancement 
with respect to $B\to\pi K$ decays~\cite{pdg}, both of which
are expected to be dominated by $b\to s$ penguin diagrams.
We can use flavor SU(3) to relate $b\to s$ and $b\to d$ penguins;
the measured branching fractions indicate that
the relative penguin contributions in the $B\to\rho\rho$ 
decays are smaller than in the $B\to\pi\pi$ case.

A more quantitative estimate of penguin contributions 
in $B\to\rho\rho$ decays can be obtained using isospin 
relations and measurements of $B\to\rho^0\rho^0$, 
$\rho^+\rho^-$, and $\rho^+\rho^0$ branching fractions 
and polarization~\cite{babar:rhoplrhomn,isorhorho}.
Since the tree contribution to the $B^0\to\rho^0\rho^0$ 
decay is color-suppressed (see Table~\ref{tab:results3}), 
the decay rate is sensitive to the penguin diagram 
and its tight experimental limit
provides tight constraints on penguin pollution.
This makes $B^0\to\rho^+\rho^-$ an ideal channel 
for the time-dependent measurement of the CKM angle 
$\alpha$. It is interesting to note that 
$B^0\to\omega\rho^0$ measurement provides comparable 
constraint on penguin contribution, but additional 
assumptions are required. 

\section*{Acknowledgments}
I am grateful for the excellent work by $\babar$ 
members who contributed results and 
made this work possible. I would like to thank 
Bob Cahn, Alex Kagan, Zoltan Ligeti, David London, 
Jim Smith, Mahiko Suzuki, and Arkady Vainshtein 
for useful discussions.

\end{document}